\documentstyle[twoside]{article}

\catcode`\@=11
\long\def\@makefntext#1{
\protect\noindent \hbox to 3.2pt {\hskip-.9pt
$^{{\eightrm\@thefnmark}}$\hfil}#1\hfill}               %CAN BE USED

\def\@makefnmark{\hbox to 0pt{$^{\@thefnmark}$\hss}}    %ORIGINAL

\def\ps@myheadings{\let\@mkboth\@gobbletwo
\def\@oddhead{\hbox{}
\rightmark\hfil\eightrm\thepage}
\def\@oddfoot{}\def\@evenhead{\eightrm\thepage\hfil
\leftmark\hbox{}}\def\@evenfoot{}
\def\sectionmark##1{}\def\subsectionmark##1{}}

\oddsidemargin=\evensidemargin
\newcounter{sectionc}\newcounter{subsectionc}\newcounter{subsubsectionc}

\renewcommand{\section}[1] {\vspace{12pt}\addtocounter{sectionc}{1}
\setcounter{subsectionc}{0}\setcounter{subsubsectionc}{0}\noindent
        {\tenbf\thesectionc. #1}\par\vspace{5pt}}
\renewcommand{\subsection}[1]
{\vspace{12pt}\addtocounter{subsectionc}{1}
      \setcounter{subsubsectionc}{0}\noindent
      {\bf\thesectionc.\thesubsectionc.{\kern1pt \bfit
#1}}\par\vspace{5pt}}
\renewcommand{\subsubsection}[1]
      {\vspace{12pt}\addtocounter{subsubsectionc}{1}
      \noindent{\tenrm\thesectionc.\thesubsectionc.\thesubsubsectionc.
      {\kern1pt \tenit #1}}\par\vspace{5pt}}
\newcommand{\nonumsection}[1] {\vspace{12pt}\noindent{\tenbf #1}
        \par\vspace{5pt}}

\newcounter{appendixc}
\newcounter{subappendixc}[appendixc]
\newcounter{subsubappendixc}[subappendixc]
\renewcommand{\thesubappendixc}{\Alph{appendixc}.\arabic{subappendixc}}
\renewcommand{\thesubsubappendixc} {\Alph{appendixc}.\arabic{subappendixc}.\arabic{subsubappendixc}}

\renewcommand{\appendix}[1] {\vspace{12pt}
        \refstepcounter{appendixc}
        \setcounter{figure}{0}
        \setcounter{table}{0}
        \setcounter{lemma}{0}
        \setcounter{theorem}{0}
        \setcounter{corollary}{0}
        \setcounter{definition}{0}
        \setcounter{equation}{0}
        \renewcommand{\thefigure}{\Alph{appendixc}.\arabic{figure}}
        \renewcommand{\thetable}{\Alph{appendixc}.\arabic{table}}
        \renewcommand{\theappendixc}{\Alph{appendixc}}
        \renewcommand{\thelemma}{\Alph{appendixc}.\arabic{lemma}}
        \renewcommand{\thetheorem}{\Alph{appendixc}.\arabic{theorem}}

\renewcommand{\thedefinition}{\Alph{appendixc}.\arabic{definition}}

\renewcommand{\thecorollary}{\Alph{appendixc}.\arabic{corollary}}
        \renewcommand{\theequation}{\Alph{appendixc}.\arabic{equation}}
        \noindent{\tenbf Appendix \theappendixc #1}\par\vspace{5pt}}
\newcommand{\subappendix}[1] {\vspace{12pt}
        \refstepcounter{subappendixc}
        \noindent{\bf Appendix \thesubappendixc. {\kern1pt \bfit #1}}
        \par\vspace{5pt}}
\newcommand{\subsubappendix}[1] {\vspace{12pt}
        \refstepcounter{subsubappendixc}
        \noindent{\rm Appendix \thesubsubappendixc. {\kern1pt \tenit
#1}}
        \par\vspace{5pt}}

\newcommand{\smalllineskip}{\baselineskip=10pt}

\def\eightcirc{
\begin{picture}(0,0)
\put(4.4,1.8){\circle{6.5}}
\end{picture}}
\def\eightcopyright{\eightcirc\kern2.7pt\hbox{\eightrm c}}

\def\abstracts#1#2#3{{

\centering{\begin{minipage}{4.5in}\baselineskip=10pt\footnotesize
        \parindent=0pt #1\par
        \parindent=15pt #2\par
        \parindent=15pt #3
        \end{minipage}}\par}}

\newcounter{itemlistc}
\newcounter{romanlistc}
\newcounter{alphlistc}
\newcounter{arabiclistc}

\newcommand{\fcaption}[1]{
        \refstepcounter{figure}
        \setbox\@tempboxa = \hbox{\footnotesize Fig.~\thefigure. #1}
        \ifdim \wd\@tempboxa > 5in
           {\begin{center}
        \parbox{5in}{\footnotesize\smalllineskip Fig.~\thefigure. #1}
            \end{center}}
        \else
             {\begin{center}
             {\footnotesize Fig.~\thefigure. #1}
              \end{center}}
        \fi}

\newcommand{\tcaption}[1]{
        \refstepcounter{table}
        \setbox\@tempboxa = \hbox{\footnotesize Table~\thetable. #1}
        \ifdim \wd\@tempboxa > 5in
           {\begin{center}
        \parbox{5in}{\footnotesize\smalllineskip Table~\thetable. #1}
            \end{center}}
        \else
             {\begin{center}
             {\footnotesize Table~\thetable. #1}
              \end{center}}
        \fi}

\def\@citex[#1]#2{\if@filesw\immediate\write\@auxout
        {\string\citation{#2}}\fi
\def\@citea{}\@cite{\@for\@citeb:=#2\do
        {\@citea\def\@citea{,}\@ifundefined
        {b@\@citeb}{{\bf ?}\@warning
        {Citation `\@citeb' on page \thepage \space undefined}}
        {\csname b@\@citeb\endcsname}}}{#1}}

\newif\if@cghi
\def\cite{\@cghitrue\@ifnextchar [{\@tempswatrue
        \@citex}{\@tempswafalse\@citex[]}}
\def\citelow{\@cghifalse\@ifnextchar [{\@tempswatrue
        \@citex}{\@tempswafalse\@citex[]}}
\def\@cite#1#2{{$\null^{#1}$\if@tempswa\typeout
        {IJCGA warning: optional citation argument
        ignored: `#2'} \fi}}

\def\@refcitex[#1]#2{\if@filesw\immediate\write\@auxout
        {\string\citation{#2}}\fi
\def\@citea{}\@refcite{\@for\@citeb:=#2\do
        {\@citea\def\@citea{, }\@ifundefined
        {b@\@citeb}{{\bf ?}\@warning
        {Citation `\@citeb' on page \thepage \space undefined}}
        \hbox{\csname b@\@citeb\endcsname}}}{#1}}

\def\@refcite#1#2{{#1\if@tempswa\typeout
        {IJCGA warning: optional citation argument
        ignored: `#2'} \fi}}

\def\refcite{\@ifnextchar[{\@tempswatrue
        \@refcitex}{\@tempswafalse\@refcitex[]}}

\def\pmb#1{\setbox0=\hbox{#1}
        \kern-.025em\copy0\kern-\wd0
        \kern.05em\copy0\kern-\wd0
        \kern-.025em\raise.0433em\box0}

\def\fnt#1#2{\footnotetext{\kern-.3em
        {$^{\mbox{\scriptsize #1}}$}{#2}}}

\def\fpage#1{\begingroup
\voffset=.3in
\thispagestyle{empty}\begin{table}[b]\centerline{\footnotesize #1}
        \end{table}\endgroup}

\def\runninghead#1#2{\pagestyle{myheadings}
\markboth{{\protect\footnotesize\it{\quad #1}}\hfill}
{\hfill{\protect\footnotesize\it{#2\quad}}}}
\font\tenrm=cmr10
\font\tenit=cmti10
\font\tenbf=cmbx10
\font\bfit=cmbxti10 at 10pt
\font\ninerm=cmr9

\font\eightrm=cmr8

\def\qed{\hbox{${\vcenter{\vbox{                      %HOLLOW SQUARE
   \hrule height 0.4pt\hbox{\vrule width 0.4pt height 6pt
   \kern5pt\vrule width 0.4pt}\hrule height 0.4pt}}}$}}

\begin{document}
\runninghead{G. K\"albermann et al.}
{Nearness through an $\ldots$}

%\normalsize\textlineskip
\thispagestyle{empty}\setcounter{page}{1}
\vspace*{0.88truein}
\fpage{1}

\centerline{\bf NEARNESS THROUGH AN EXTRA DIMENSION}
\vspace*{0.035truein}

\vspace*{0.37truein}
\centerline{\footnotesize G. K\"albermann$^1$ \& H. Halevi$^2$}

\centerline{\footnotesize \it
$^1$Faculty of Agriculture and Racah Institute of Physics,}
\baselineskip=10pt
\centerline{\footnotesize \it
Hebrew University, Jerusalem 91904, Israel.(permanent address)}
\baselineskip=10pt
\centerline{\footnotesize \it
Cyclotron Institute, Texas A\&M University College Station, TX 77843, USA}
\baselineskip=10pt
\centerline{\footnotesize \it
$^2$4 Manee St., Jerusalem, Israel}

%\date{\today}

\baselineskip 5mm

\vspace*{0.21truein}

\abstracts{
It is shown that if our visible universe is a thin trapped shell
in a five-dimensional universe,
all matter in it may be connected almost instantaneously through
the fifth dimension. What appears to be action at a distance
is then understood as undetectable ultrafast communication.}{}{}

%\pacs{PACS numbers: 03.50.-z, 03.50.De}

$$$$
\section{The universe as a shell}

Several authors in the physics literature speculated about the possibility
that our universe is a thin shell in a large dimensional hyper-universe.

Wesson [1], tried to explain the very existence of mass and charge
of elementary particles as originating from the properties of a universe
with extra dimensions. He found that mass could be related to the fifth
coordinate and charge to the momentum along it.

Visser [2] before him showed that the trapping of our universe can be
implemented mathematically by using a large cosmological constant.
Squires [3] later implemented the trapping by using a scalar
field that becomes effectively confined to one dimension less than the
original space, again, by using a cosmological constant.
The trapped universe is essentially flat.
 
More formal arguments for multidimensional spaces were given by Beciu [4].

These proposals are an alternative to the usual claim that extra dimensions
are curled up to an unobservable size, an assertion that is not devoid of
problems.[5]

In the alternative approach of refs.[1-4], 
our universe becomes a thin shell in a larger hyper-universe.

The trapping has to be gravitationally repulsive in nature.
A cosmological constant does indeed produce the effect, however
this is not the only possibility. If the inner-outer space to the shell
is filled with negative energy, then it can trap such a shell effectively.

In the present work we will show that, regardless of the mechanism of
trapping,
its very existence implies that all matter in the universe may
be connected through the fifth dimension by means of electromagnetic
(or other) signals in an undetectable amount of time.

All matter and energy in the universe is then tied up together in a
manner that usual locality and causality on the shell would prohibit.
This may in turn explain not only phenomena related to quantum mechanical
behavior, but also macroscopic action at a distance.

A grotesque way to view the universe would be comparing the matter in it
to puppets that are connected by cables to a puppeteer that holds 
all the strings together. A perturbation in the hands of the puppeteer
causes a turmoil in all the other strings in unison.

In the next section a simple model will investigate these assertions and
briefly comment about its implications.

The model addresses what appears to be action at a distance
in terms of ultrafast communication. Action at a distance is, in our
opinion, a mirage, the real situation is an incessant bombardment by
ultrafast radiation that fills the universe's shell.

\section{Nearness through the fifth dimension}

Consider a five-dimensional metric space with line element

\begin{equation}\label{element}
ds^2~= F(R,t)~dt^2-G(R,t)~dR^2-c(R,t)~d{\Omega_3}^2
\end{equation}

where $F$ and $G$ are for simplicity taken as depending only on the
fifth coordinate and the cosmic time $t$ and, 
$d{\Omega}_3$ is the volume element 
of the three dimensional subspace on the surface.

The above ansatz corresponds
to a hyperspherically symmetric solution of five dimensional Einstein
gravity, provided Einstein's gravity still works
in the hyper-universe, as one might hope. We still regard 
space-time inside the shell as a Riemannian manifold.

Define the center of the hypersurface by $R_0$, such that
$F(R_0,t)=1,~c(R_0,t)=S(t)$, with $S$ the scale parameter of our expanding
universe. This coordinate choice 
defines what is understood by the cosmic time on the shell, at $x=0$.

This is similar to the De-Sitter picture for which the
intrinsic radius of the universe is merely the radius of
curvature of the hypersurface embedded in the larger flat universe.
The differences arise only from the fact that we do not constrain
the shell to be of zero thickness as for the De-Sitter case, or
the cases investigated in the past.[1,4] 
The shell has a finite -albeit minuscule- width.

We now simplify the problem a bit more by taking
all the metric functions to be time independent, namely we choose
$t=t_0$, the present epoch.
We do so because the scale function of conventional Friedmann-Robertson-Walker
expansion is the inverse Hubble constant of the order of the radius of
the universe, extremely large
as compared to the supposed thickness of the shell.
Time evolution of the shell 
is not important for the time scales that will be involved in the present
investigation.

We have chosen the signature of the extra dimension as space-like
inspired in previous results.[1,2]
However, as we will
show later, the signature may change at the inner-outer
boundaries of the shell.

Regardless of the trapping mechanism (cosmological constant, negative
energy)
the functions $F$ and $G$ may be expanded around the equilibrium (or
quasiequilibrium, due to the time dependence) radius $R_0$.
At fixed time the expansion has to be a constant and a quadratic term
in the distance $x=R-R_0$. The linear term has to vanish due to the
equilibrium condition, otherwise, all the mass in the universe would suffer
a constant force that will eventually drag it to a new equilibrium position
and we can always call this new position $R_0$. 

In any event, we will
assume, that the extension of the shell into the fifth dimension is
so tiny as compared to any reasonable scale length -as it should be
if its effect remains undetected in the dynamics of bodies-, that
our approximation will be valid. As mentioned above, one way to actually
obtain the expansion is to consider a positive mass shell
in a cosmological constant filled universe as in ref.[4].

Moreover, as in any outer (to the inner-outer universe) 
Schwartzschild-like solution, we will take $F=G^{-1}$.[4]

Hence we have

\begin{equation}\label{F}
F(R,t)|_{R_0}=~A+B~x^2
\end{equation}

where $A$ and $B$ are constants. 

In order to fulfill the constraint of a flat space-time on the shell
at our present time, we redefine our coordinates to have $A=1$[6].

We will further make the assumption that the trapping is due to
the repulsion from the inner and outer regions, hence $B$ has
to be positive, for the particles to be repelled into the shell.
(Recall the relation between gravitational potential and metric
in the Newtonian approximation).

Collapse is prevented by the pressure of the mass in the shell, although
space where mass is sparse may be penetrated by the matter of the
inner and outer regions, producing important effects.
One such effect would be similar to that provided by the hypothetical
dark matter. 

Instead of having more matter inside the galaxies that is unseen, there
could be unseen negative mass outside them. Also, if the
shell is cracked suddenly and mass from the inner and outer regions
coalesces with the mass in the shell they could annihilate
producing enormous bursts of energy, an alternative to
black hole generated active galactic nuclei.

Hence
\begin{equation}\label{fnew}
F~=~G^{-1}= 1+k~x^2
\end{equation}

with $k>0$.

In order to find out the functional dependence of $c(x=R-R_0)$, we 
insert the metric of eq.(1), with the ansatz of eq.(3) for
the metric functions $F$, and $G$ in the five dimensional Einstein tensor.

It is found that the function 
$c(x)\approx~e^{-\beta~x^2/2}$ yields, for small $x$

\begin{eqnarray}\label{tensor}
G_{00}=-\frac{3}{2}~\beta
\end{eqnarray}

All the other components of the Einstein tensor vanish provided that $\beta=k$.
This is not the most general solution, but, it will suffice for the present
purposes.

Inserting eq.(4) in the Einstein equations
with a cosmological term and an ideal fluid inside the shell,
we find that the energy momentum tensor in the
comoving frame of the shell can be chosen to have vanishing energy density 
and a pressure $p\approx\frac{3~k}{16\pi~G}$ provided by the
cosmological constant, whith $G$ the 
gravitational constant. 
This solution is only one amongst many other choices.
Alternatively we could have chosen
a solution with no cosmological constant, no pressure, and an enormous
energy density in the shell, or any solution in between. Due to the fact
that our universe seems to be essentially vacuous, we favor the first 
solution.

If we choose the shell thickness to be microscopical
in size, the value of the cosmological constant will be enormous $\lambda=
3/4~k$.

The cosmological constant prevents
the collapse of the shell against the gravitational squeeze
produced by the harmonic potential.
The effect of the constant
is to maintain the shell in equilibrium against
the repulsion of the inner-outer regions. 
The need for a large cosmological constant is 
consistent with previous works[2].

The constant $k$ should be large enough, for the
thickness of the shell to be smaller than the radii of nuclei,
smaller than the deep inelastic scattering scale of the experiments
carried until now, otherwise, 
its imprint should have become visible by now.
\footnote{It would be intriguing to find out about the
influence of the thickness of the shell on electromagnetic
processes in a Kaluza-Klein type
of approach, without curled up extra dimensions}

From eqs.(1,3), it appears that, space-time will change
signature at the boundaries of the shell. The inner-outer universe
is then shielded from the shell by a horizon, at least classically.
Quantum effects, like Hawking radiation might still be able to
cross the boundary, as for the case of a black hole.

Let us now consider signal propagation
along a path that goes in the $R$ direction, then
across it and back.

Our measure of time on the shell is the cosmic time $t$,
and we refer every process to it.
Suppose a signal whose speed is the velocity of light $c=1$ in our units,
starts traveling from $x=0$, the center of the shell to some fixed x
inside it, then travels a distance $L$ perpendicular to $R$ at
fixed $x$ and returns from $x$ to $x=0$ back to a
point at a coordinate distance $L$ from the initial point. 
For such
a scenario to occur, radiation has to propagate in a direction that
is not completely perpendicular to the hypersurface, scatter
inside it, or be reflected at some hypothetical edge. 
The tangential component of the velocity, can provide the
motion along the surface. Reflection and scattering
can modify the frequency of the radiation, but, it seems
unlikely the time lapse of the trip will be affected by these
processes.
It is then sufficient to consider the path chosen.

In principle, zero mass radiation can
reach any value of x, even large values as compared to $\sqrt{1/k}$. 
The farther out radiation travels, the larger the effect. 

Using three dimensional spherical
coordinates, $L$ along the radial distance in three-dimensional
space, and for fixed angles, we find

\begin{equation}\label{elem}
ds^2=(1+kx^2)dt^2-\frac{1}{(1+k~x^2)}~dR^2-e^{-k~x^2/2}~dl^2
\end{equation}

The definition of $dl$ is such 
that at $x=0$ space-time is essentially Minkowski, as observed.

Further, for radiation we still have $ds^2=0$.[6]

Our choice of $k$ will be of the order of $R_{GUT}^{-2}$, where
$R_{GUT}\approx10^{-31}m$ stands for the grand unified theories scale. 
We do this because we want to encompass democratically all the known
interactions, and, in order to avoid quantum gravity
effects that will enter at much smaller scales of the order
of $R_{Planck}\approx~10^{-35}m$.

However, any microscopic scale will be a viable choice. We take this
value for the sake of exemplification

Hence

\begin{equation}\label{t}
t=~2\int_0^x{\frac{dx}{1+k~x^2}}~+~\int_0^L{\frac{dl~e^{-k~x^2/4}}
{\sqrt{1+kx^2}}}
\approx~\frac{L~e^{-k~x^2/4}}{\sqrt{1+kx^2}}
\end{equation}

Where we have used $x=R-R_0$ and neglected the first integral
because it is of the order of $t\approx10^{-39}sec$.

If the signal climbs up the harmonic potential and back far enough
in $x$, the coordinate time becomes negligible.

With $x= 15~R_{GUT}\approx 1.5~10^{-30}m$, several times
the radius of curvature of the shell, and $L= 100 Mpc$, 
the time taken by radiation to traverse this cosmic distance is
$t=2.5~10^{-10}sec$. A ridiculously small time 
as compared to the 326 Million years 
needed for the light to traverse this distance along the direction 
perpendicular to R on the surface. 
Recall that on the surface, $x=0$.

Due to the crudeness of the approximations used in order
to derive the above result, we should not attach too much rigor to
the actual numbers. The effect is, nevertheless, evident.

This amazing mechanism might be at work for all the processes we recognize
as action at a distance. Gravity waves are not an exception.
However, we are still far from explaining in mathematical detail
how will this actually generate static instantaneous-like
interactions between far away bodies, that could not be communicated
causaly through the shell. It is, nevertheless a line of thought
that has not been explored before and deserves further attention.

The scheme can fail if, either radiation is not able to
climb up the potential very far uphill, because
the harmonic is an extremely crude approximation and other terms may come
into play creating perhaps a horizon, or, if it is
absorbed in the inner and outer spaces.
Both options seem unlikely due to the repulsive nature of the
exterior-interior spaces.

The signal becomes red shifted and blue shifted enormously, but will arrive
at its destination with the same frequency as emitted.

All the radiation that is emitted in the $R$ direction is then
detected in no-time by all the other particles in the universe at random.
This radiation will oscillate back and forth in the R direction and most
certainly fill the shell completely. This wandering radiation is
a poor man's model of what one would call an action at a distance.
The fact that we do not violate causality and locality is because
both are distorted enormously by the potential of the shell.
This in turn might have some bearing to the nonlocality witnessed
in quantum mechanics. Instead of having alternative hidden-variable
theories we could think about hidden-dimension theories.

In summary, if our universe is a thin shell immersed in a higher
dimensional hyperuniverse, and if the shell is trapped by some
repulsive force or analogous mechanism, then what seems to be
action at a distance, such as the action of static potentials
could be due to action through the fifth dimension.

\nonumsection{References}
%begin{thebibliography}{99}
\begin{itemize}
\item[{1.}] P. S. Wesson and H. Liu, Int. Jour. of Theor. Phys. {\bf
36}, 1865 (1997) and references therein.

\item[{2.}] M. Visser Phys. Lett. {\bf B159} ,22 (1985).

\item[{3.}] E. J. Squires, Phys. Lett. {\bf B167}, 286 (1985).

\item[{4.}] M. I. Beciu, Europhys. Lett. {\bf 12}, 229 (1990).

\item[{5.}] M. J. Duff, {\sl Supersymmetry, Supergravity and related
topics}, World Scientific, Singapore, 1984.

\item[{6.}] S. Weinberg, {\sl Gravitation and Cosmology} J. Wiley
and Sons, 1972.
\end{itemize}
%\end{thebibliography}

\end{document}